\documentclass[
aps,prd,
preprintnumbers,
amsmath,
amssymb,
nofootinbib,
preprintnumbers
]{revtex4}
\usepackage{here}
\usepackage{footnote}
\usepackage{comment,braket}
\usepackage{epsf}
\usepackage{amsmath}
\usepackage{graphics}
\usepackage{amsfonts}
\usepackage{amssymb}
\usepackage{latexsym}
\usepackage{color}
\usepackage{natbib}
\usepackage{graphicx}
\usepackage{hyperref}
\usepackage{array,multirow}

\usepackage[svgnames]{xcolor}
\definecolor{phthaloblue}{rgb}{0.0, 0.06, 0.54}
\hypersetup{
    colorlinks=true,
    linkcolor=blue,
    citecolor=blue,
    filecolor=blue,
    urlcolor=blue,
    }

\begin{document}
\title{Greybody Factors Imprinted on Black Hole Ringdowns:\\
an alternative to superposed quasi-normal modes}
\author{Naritaka Oshita}
\affiliation{Center for Gravitational Physics and Quantum Information (CGPQI),
Yukawa Institute for Theoretical Physics (YITP), Kyoto University, 606-8502, Kyoto, Japan}
\affiliation{The Hakubi Center for Advanced Research, Kyoto University,
Yoshida Ushinomiyacho, Sakyo-ku, Kyoto 606-8501, Japan}
\affiliation{RIKEN iTHEMS, Wako, Saitama, 351-0198, Japan}
\preprint{YITP-23-111}
\preprint{RIKEN-iTHEMS-Report-23}

\begin{abstract}
It is shown that the spectral amplitude of gravitational-wave (GW) ringdown of a Kerr black hole sourced by an extreme mass ratio merger can be modeled by the {\it greybody factor}, which quantifies the scattering nature of the black hole geometry. The estimation of the mass and spin of the remnant is demonstrated by fitting the greybody factor to GW data without using black hole quasi-normal modes. We propose that the ringdown modeling with the greybody factor may strengthen the test of gravity as one can avoid the possible overfitting issue and the start time problem in the ringdown modeling with superposed quasi-normal modes.
\end{abstract}

\maketitle

\section{Introduction}
The Kerr solution, describing a spinning black hole, is one of the most simplest solutions to the Einstein equation. Based on the black hole no-hair theorem \cite{Israel:1967wq,Israel:1967za,Carter:1971zc}, the spacetime structure near an astrophysical Kerr black hole is characterized by two parameters only, i.e, the mass $M$ and angular momentum $J$ of the black hole. Therefore, a black hole is a suitable site to test gravity in strong gravity regimes. In the context of the test of the no-hair theorem, the black hole spectroscopy \cite{Echeverria:1989hg,Finn:1992wt,Berti:2005ys} has been actively studied so far. The black hole spectroscopy is an extraction of each black hole quasi-normal (QN) mode \cite{Leaver:1986gd,Sun:1988tz,Andersson:1995zk,Glampedakis:2001js,Glampedakis:2003dn,Nollert:1992ifk,Andersson:1996cm,Nollert:1998ys} from gravitational wave (GW) ringdown which is a superposition of multiple QN modes. There are infinite number of QN modes and each mode has a complex frequency  $\omega = \omega_{lmn} \in {\mathbb C}$ labeled by the overtone number $n$ for each angular and azimuthal mode $(l,m)$. The real and imaginary part of $\omega_{lmn}$ represent the frequency and damping rate of the mode, respectively. 

GW ringdown appears after the inspiral phase of a binary black hole system. If the ringdown starts around the strain peak, it would be possible to measure several QN modes included in a ringdown signal by truncating GW data before the assumed start time of ringdown and by fitting several QN modes to the truncated data \cite{Giesler:2019uxc}.
However, some issues in the black hole spectroscopy with (superposed) QN modes have been pointed out like the start time problem \cite{Sun:1988tz,Nollert:1998ys,Andersson:1996cm,Berti:2006wq} and overfitting problem \cite{Baibhav:2023clw}.
Then, it would be natural to ask if there is another nice quantity being suitable to test gravity other than QN modes.

In this paper, we propose that the black hole greybody factor, $\Gamma_{lm} (\omega)$, would be an important quantity in the test of the no-hair theorem and the estimation of the remnant parameters from GW ringdown. We here consider a particle plunging into a massive black hole as a source of GW. Then we show that for $(l,m)=(2,2)$, $\Gamma_{lm}$ can be imprinted on the GW spectral amplitude $|\tilde{h}_{lm} (\omega)|$ in $\omega \gtrsim f_{lm} \equiv \text{Re} (\omega_{lm0})$ with the form of
\begin{equation}
|\tilde{h}_{lm} (\omega)| \simeq c_{lm} \times \gamma_{lm} (\omega) \equiv c_{lm} \times \sqrt{1-\Gamma_{lm} (\omega)}/\omega^3 \ \text{for} \ \omega \gtrsim f_{lm},
\label{main}
\end{equation}
where $\omega$ is a frequency of GWs and $c_{lm}$ is a constant corresponding to the GW amplitude.
The frequency dependence of the greybody factor is determined by the two remnant parameters only, i.e., the mass and spin of the black hole. It means that if (\ref{main}) holds, one can detect the greybody factor from the ringdown to test the no-hair theorem as the spectral amplitude in $\omega \gtrsim f_{lm}$ corresponds to the ringdown signal. The reflectivity ${\cal R}_{lm} \equiv 1-\Gamma_{lm}$ has an exponential damping at high frequencies ($\omega \gtrsim f_{lm}$) and the strength of the damping in the frequency domain is unique for the remnant mass and spin like a complex QN mode frequency. As the damping in ${\cal R}_{lm}$ is strong for rapid spins, the reflectivity well govern the dependence of $|\tilde{h}_{lm}|$ on frequency and our model works well especially for rapidly spinning remnant black holes.
One of the important difference in the ringdown modeling with the greybody factor and QN modes is that given the remnant mass $M$ and spin $j (\equiv J/M^2)$, we know where the universal damping of ${\cal R}_{lm} = 1-\Gamma_{lm}$ appears in the frequency space, i.e., $\omega \gtrsim f_{lm} (M,j)$, but it is unknown when the excitation of superposed QN modes appears in the time domain, which is recognized as the start time problem or the time-shift problem.

The original idea of the modeling of ringdown with the greybody factor was introduced in the previous paper by the author \cite{Oshita:2022pkc}. In this paper, we investigate the importance of the greybody factors in the ringdown sourced by an extreme mass ratio merger in more detail. In Sec. \ref{sec_GWwaveform}, we explain our methodology to compute GW waveform in the linear perturbation regime. The definition and the property of the greybody factor is provided in Sec. \ref{sec_greybody}. In Sec. \ref{sec_why}, we study why the greybody factor can be imprinted on the GW ringdown by carefully considering the effect of the source term. In Sec. \ref{sec_exp_decay}, we investigate the exponential damping in the greybody factor and how it is consistent with the exponential damping in the spectral amplitude of the GW ringdown. In Sec. \ref{sec_mismatch}, we perform the measurement of the remnant mass and spin only with the fit of the greybody factor. In Sec. \ref{sec_conclusion}, our conclusion is provided and we discuss the pros and cons of using the greybody factor and QN modes in the test of the no-hair theorem, measurement of the remnant quantities, and the modeling of GW ringdown. Throughout the manuscript, we use the natural unit of $c=\hbar=1$ and $G=1$.

\section{Formalism}
\label{sec_formalism}
In this section, we describe how we compute GW spectral amplitude for an extreme mass ratio merger and the greybody factor of a spinning black hole. Here we concentrate on a particle plunging into the hole and its trajectory is restricted on the equatorial plane.
\subsection{extreme mass ratio merger and gravitational waveform}
\label{sec_GWwaveform}
The background geometry is approximated by the Kerr spacetime when we consider an extreme mass ratio merger with a massive black hole. Therefore, the background geometry can be covered by the Boyer-Lindquist coordinates $(t,r,\theta,\phi)$ and one can compute the GW spectrum $\tilde{h}_{lm} (\omega)$ sourced by the merger event in a linear manner. Let us begin with solving the Sasaki-Nakamura equation \cite{Sasaki:1981sx}:
\begin{equation}
\left[ \frac{d^2}{dr^{\ast}{}^2} -F_{lm} \frac{d}{dr^{\ast}} - U_{lm} \right] X_{lm} = \rho_{lm},
\label{SN_equation}
\end{equation}
where the explicit forms of $F_{lm}$ and $U_{lm}$ are given in the original paper by Sasaki and Nakamura \cite{Sasaki:1981sx}, and the spectrum $\tilde{h}_{lm}$ is obtained from the perturbation variable $X_{lm}$ as is shown later explicitly. The source term $\rho_{lm}$ depends on the plunging orbit of a particle with mass $\mu$. The form of $\rho_{lm}$ for the plunging particle on the equatorial plane ($\theta = \pi/2$) is \cite{Kojima:1984cj}
\begin{equation}
\rho_{lm} = \frac{\gamma_0 \Delta}{(r^2+a^2)^{3/2} r^2} W \exp{\left(-i \int^r \frac{K(r')}{\Delta(r')} dr' \right)},
\label{source_e}
\end{equation}
where $\Delta (r) \equiv r^2 -2Ma +a^2$, $K (r) \equiv (r^2+a^2) \omega - am$, and $a \equiv J/M$. The functions $\gamma_0$ and $W$ are shown in Appendix A and B in Ref. \cite{Kojima:1984cj}, respectively.
\begin{figure}[b]
\centering
\includegraphics[width=0.6\linewidth]{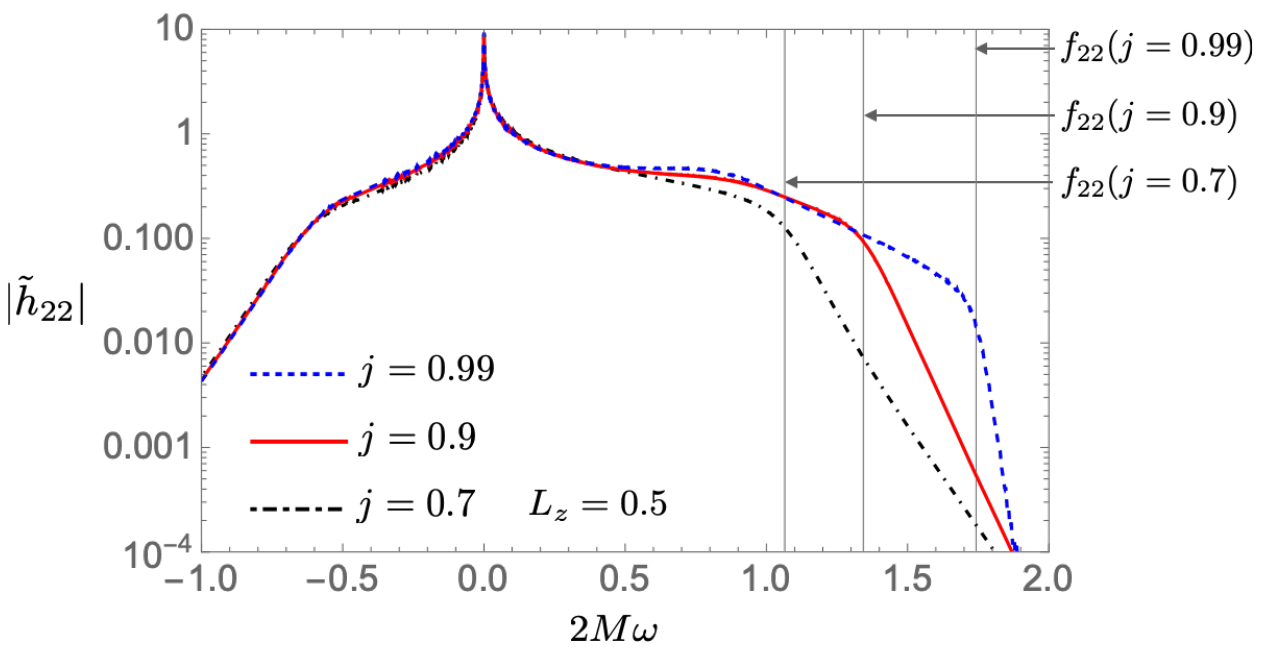}
\caption{Spectral amplitude $|\tilde{h}_{22}|$ for $j=0.7$ (black dash-dotted), $0.9$ (red solid), and $0.99$ (blue dashed). The source term is obtained with $L_z = 0.5$ and the observation angle is $\theta = \pi/2$. The black vertical lines show the real part of the fundamental QN mode frequency $f_{22} (j)$. The exponential damping of the spectral amplitudes appear in $\omega \gtrsim f_{22}$.}
\label{pig_spec_amp}
\end{figure}
The trajectory of a particle is determined by the following differential equations \cite{Carter:1968rr,Kojima:1984cj}:
\begin{align}
r^2 \frac{dt}{d\tau} &= -a (a-L_z) + \frac{r^2+a^2}{\Delta} P,\\
r^2 \frac{d\phi}{d\tau} &= -a (a-L_z) + \frac{a}{\Delta} P,\\
r^2 \frac{dr}{d\tau} &= -\sqrt{R},\\
\theta &= \pi/2,
\end{align}
where $P\equiv r^2+a^2-L_z a$, $R\equiv 2M r^3-L_z^2 r^2 + 2Mr(L_z-a)^2$, $\mu L_z$ is the orbital angular momentum, and $\tau$ is the proper time of the particle. We obtain the source term $\rho_{lm}$ by substituting the trajectory of the particle, $(t(\tau), r(\tau), \theta = \pi/2, \phi (\tau))$, into (\ref{source_e}).
We then numerically compute the Sasaki-Nakamura equation with the source term for a plunging orbit on the equatorial plane $\rho_{lm}$. The GW spectral amplitude we obtained for $l=m=2$ are shown in Figure \ref{pig_spec_amp}.

\subsection{greybody factors}
\label{sec_greybody}
\begin{figure}[t]
\centering
\includegraphics[width=0.65\linewidth]{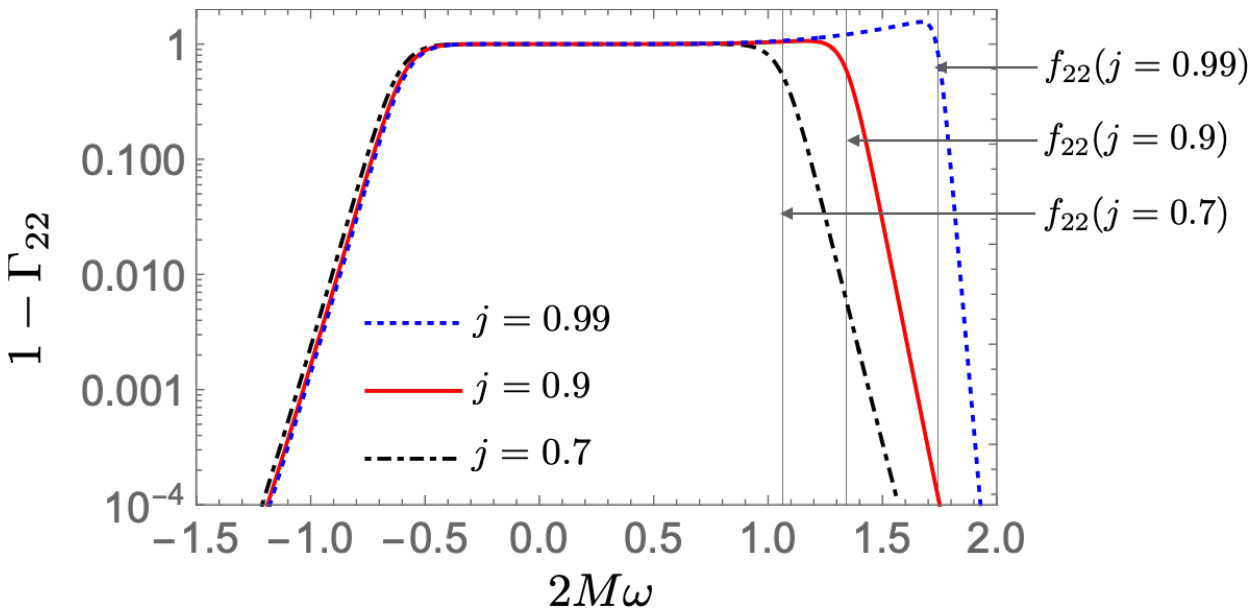}
\caption{Reflectivity ${\cal R}_{lm} = 1-\Gamma_{lm}$ for $(l,m)=(2,2)$ with $j=0.7$ (black dot-dashed), $0.9$ (red solid), and $0.99$ (blue dashed). The black vertical lines show the real part of the fundamental QN mode frequency $\omega = f_{22}$. The exponential damping of the reflectivity appear in $\omega \gtrsim f_{22}$.}
\label{pig_greybody_factor}
\end{figure}

\begin{table}
\centering
\setlength{\tabcolsep}{8pt}
\begin{tabular}{c||ccccccc}
 \hline
 spin parameter $(j)$ & 0.001& 0.1& 0.2& 0.3 & 0.4& 0.5 & 0.6\\
 \hline
 decay frequency $(T_{22})$   & 0.067    &0.066 &   0.065 &   0.064 &   0.063 &   0.062& 0.060\\
 \hline
\end{tabular}
\end{table}
\begin{table}
\centering
\setlength{\tabcolsep}{8pt}
\begin{tabular}{c||ccccccc}
 \hline
 spin parameter $(j)$ & 0.7 & 0.8 & 0.9 & 0.95 & 0.99 & 0.995 & 0.998\\
 \hline
 decay frequency $(T_{22})$    &0.057 &   0.053 &   0.045 &   0.036 &   0.019&   0.014 & 0.0096\\
 \hline
\end{tabular}
\caption{The value of $T_{lm}$ $(l=m=2)$ with respect to the spin parameter $j$. We read the value of $T_{22}$ from $1-\Gamma_{22} (\omega)$ in the range of $\alpha \times f_{22} \leq \omega \leq 1.99/(2M)$ where we choose a constant $\alpha$ in the range of $1 \leq \alpha \leq 1.2$. The constant $\alpha$ should be larger for a lower spin in order for $1- \Gamma_{22}$ to be well approximated with $e^{-(\omega- f_{22})/T_{22}}$ in the frequency range. The value of $\alpha$ we set is shown in (\ref{greybody_reading}) in Appendix \ref{app_numerical_metho}.}
\label{table_tlm}
\end{table}
The greybody factor quantifies the absorptive nature of a black hole geometry and is independent of the source term. It is determined only by the no-hair parameters of a black hole, i.e., the mass and spin of a black hole, like the black hole quasi-normal modes. We obtain the greybody factor by computing a homogeneous solution to the Sasaki-Nakamura equation $X_{lm} = X_{lm}^{\rm (hom)}$ with the boundary condition of
\begin{equation}
X_{lm}^{\rm (hom)} = e^{-i k_{\rm H}r^{\ast}} \ \text{for} \ r^{\ast} \to - \infty,
\end{equation}
where $k_{\rm H} \equiv \omega - m \Omega_{\rm H}$ with $\Omega_{\rm H} \equiv j/(2r_+)$.
We then read the asymptotic ingoing and outgoing amplitudes at a distant region as
\begin{equation}
X_{lm}^{\rm (hom)} = A_{\rm in} e^{-i \omega r^{\ast}} + A_{\rm out} e^{i \omega r^{\ast}} \ \text{for} \ r^{\ast} \to \infty.
\end{equation}
The reflectivity of the angular momentum barrier is given by \cite{Brito:2015oca,Nakano:2017fvh}
\begin{equation}
{\cal R}_{lm} \equiv \left|\frac{C}{c_0}\right|^2 \left|\frac{A_{\rm out}}{A_{\rm in}} \right|^2 \equiv 1-\Gamma_{lm},
\end{equation}
where ${\cal R}_{lm}$ and $\Gamma_{lm}$ are the reflectivity and the greybody factor (i.e., transmissivity), respectively. The factors $|C|^2$ and $c_0$ are \cite{Starobinsky:1973aij,Starobinsky_Churilov}
\begin{align}
|C|^2 &\equiv \lambda^4 + 4 \lambda^3 + \lambda^2 (-40 a^2 \omega^2 + 40 am\omega + 4) + 48 a \lambda \omega (a \omega + m) + 144 \omega^2 (a^4 \omega^2 -2 a^3 m \omega + a^2 m^2 + M^2),\\
c_0 &\equiv \lambda (\lambda + 2) -12 a \omega (a \omega - m) -i 12 M \omega,
\end{align}
respectively, and $\lambda$ is the separation constant of the spin-weighted spheroidal harmonics. We numerically compute the greybody factor by solving the Sasaki-Nakamura equation\footnote{For more details of our numerical computation, see Appendix \ref{app_numerical_metho}. Recently, the greybody factor of Kerr black holes was analytically computed by solving the connection problem of the confluent Heun equation and its analytic form was obtained in Ref. \cite{Bonelli:2021uvf}.}. Our computation reproduce the exponential decay of ${\cal R}_{22}$ at high frequencies ($\omega \gtrsim f_{22}$) and the superradiant amplification at $\omega < m \Omega_{\rm H}$ as is shown in Figure \ref{pig_greybody_factor}. The exponential damping of ${1-\Gamma_{lm}}$ at high-frequency region ($\omega \gtrsim f_{lm}$) can be approximated as
\begin{equation}
1-\Gamma_{lm} \simeq e^{-(\omega - f_{lm})/T_{lm}},
\end{equation}
where $T_{lm}$ quantifies the strength of the exponential damping of ${\cal R}_{lm}$ in the frequency domain.
The factor $T_{lm}$ is a {\it no-hair} quantity which depends only on the mass and spin of the remnant black hole like the QN mode frequency. The values of $T_{lm}$ extracted from our numerical data are shown in Table \ref{table_tlm} and the fitting methodology we used is provided in Appendix \ref{app_numerical_metho}.

\section{Greybody Factors in Ringdown}
\begin{figure}[t]
\centering
\includegraphics[width=0.9\linewidth]{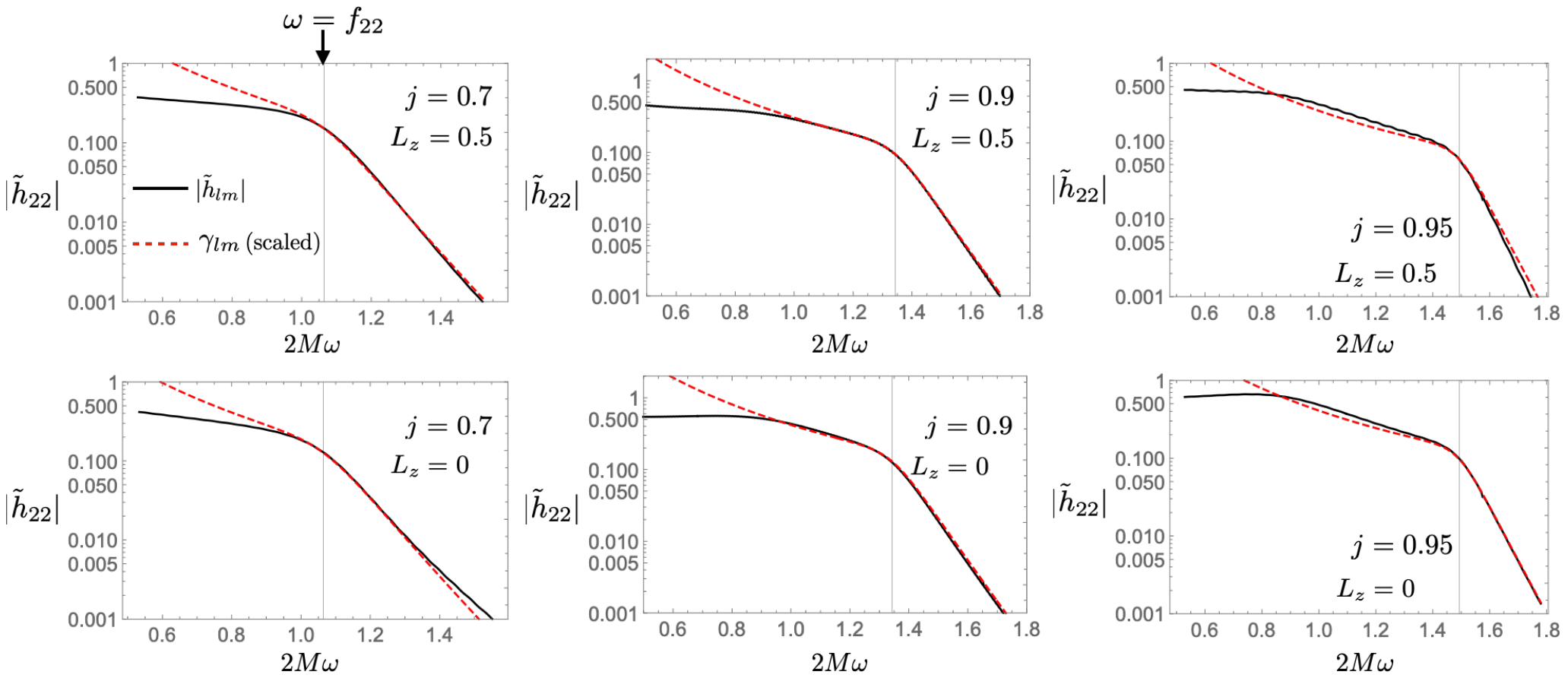}
\caption{The comparison of the GW spectral amplitude $|\tilde{h}_{22}|$ and a function $\gamma_{22} (\omega) = \sqrt{1-\Gamma_{22}}/\omega^3$ determined by the greybody factor $\Gamma_{22}$. The observation angle $\theta$ is set to $\pi/2$. For other values of $\theta$, see Appendix \ref{app_theta}. The match between the two quantities for $\omega \gtrsim f_{22}$ is explained in Sec. \ref{sec_why}. The vertical black lines indicate $\omega = f_{22}$.}
\label{pig_comp}
\end{figure}

\begin{figure}[t]
\centering
\includegraphics[width=0.9\linewidth]{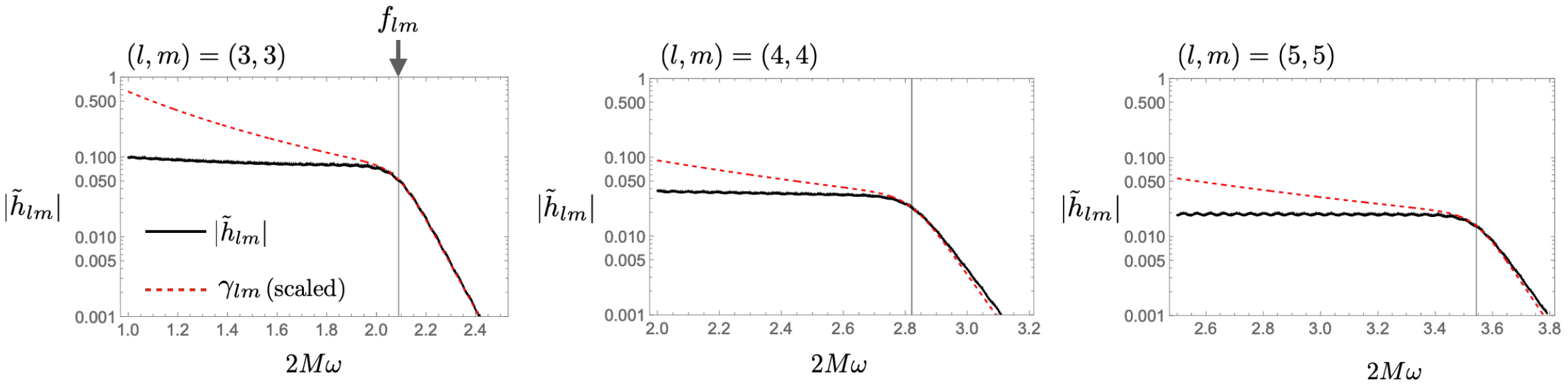}
\caption{The comparison of the GW spectral amplitude $|\tilde{h}_{lm}|$ and a function $\gamma_{lm} (\omega) = \sqrt{1-\Gamma_{lm}}/\omega^3$ for $(l,m)=(3,3)$, $(4,4)$ and $(5,5)$. The vertical black lines indicate $\omega = f_{lm}$.
We set $j=0.9$, $L_z=0.5$, and $\theta = \pi/2$.}
\label{pig_345}
\end{figure}

In this Section, we study the greybody factor imprinted on GW ringdown. As is shown in Figure \ref{pig_comp}, we find that   the greybody factor can model the spectral amplitude of GW ringdown in $\omega \gtrsim f_{lm}$ when the GW is sourced by an extreme mass ratio merger. This still holds even for some higher harmonic modes (Figure \ref{pig_345}) and for various values of the orbital angular momentum $\mu L_z$ and the spin parameter $j$ of the massive black hole. We also confirm that the frequency dependence of GW spectrum is insensitive to the observation angle $\theta$ as shown in Appendix \ref{app_theta}. This universal nature in the black hole ringdown is important to test the no-hair theorem with high precision by combining the black hole spectroscopy. 

We here study how the greybody factor can be imprinted on the ringdown signal. We also demonstrate the estimation of the remnant mass and spin by using the greybody factor only. We then find that it works well, which implies that the greybody factor is important to test the black hole no-hair theorem. Not only using the QN modes but also using the greybody factor would enhance the accuracy of the test of gravity and measurability of the remnant quantities at least for extreme mass ratio mergers.

\subsection{Why are the greybody factors imprinted on ringdown?}
\label{sec_why}
\begin{figure}[t]
\centering
\includegraphics[width=0.46\linewidth]{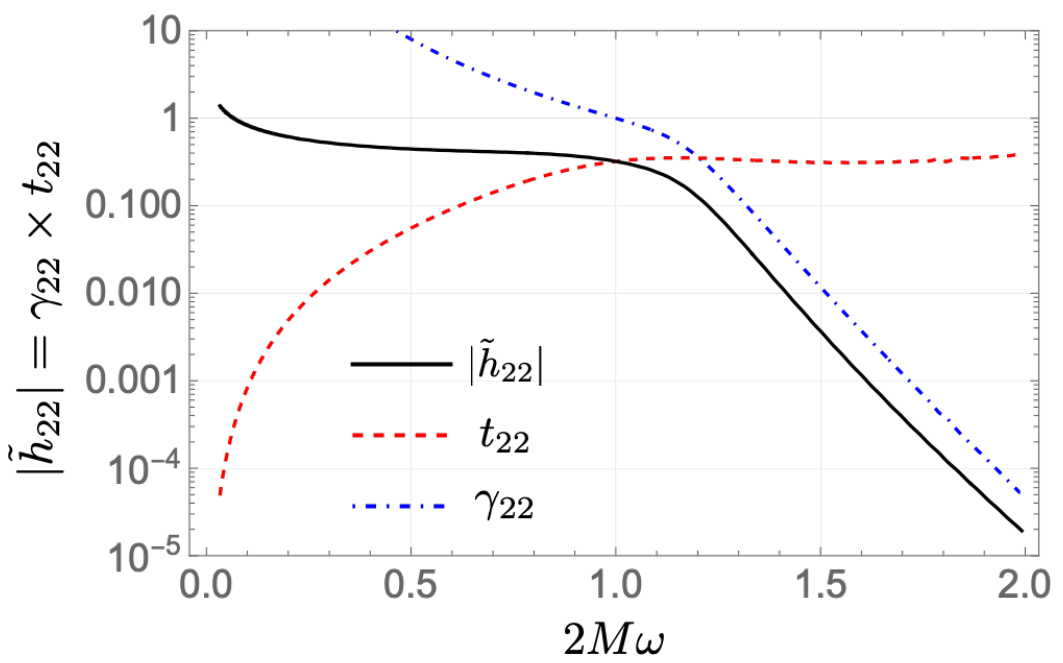}
\caption{Plot of the GW spectral amplitude $|\tilde{h}_{22}|$ (black solid), $\gamma_{22}$ (blue dot-dashed), and $t_{22}$ (red dashed) for $j=0.8$, $L_z=0$, and $\theta = \pi/2$.}
\label{pig_ref_grey}
\end{figure}

We here discuss why the greybody factors can be imprinted on the ringdown for extreme mass ratio mergers. For higher mass ratios, the background geometry is governed only by a massive black hole and the perturbation theory in the Kerr spacetime works to compute GW waveform. The GW strain, $h$, is given by 
\begin{align}
h = \sum_{l,m} \frac{e^{i m \phi}}{\sqrt{2 \pi} r} \int d\omega \tilde{h}_{lm} (\omega) e^{-i\omega t}  &= \sum_{l,m} \frac{e^{im\phi}}{\sqrt{2 \pi} r} \int d\omega \frac{-2}{\omega^2} {}_{-2} S_{lm}(a\omega, \theta) R_{lm}(\omega) e^{-i \omega t},\\
&= \sum_{l,m} \frac{e^{im\phi}}{\sqrt{2 \pi} r} \int d\omega \frac{-2}{\omega^2} {}_{-2} S_{lm}(a\omega, \theta) \frac{A_{\rm out}(\omega)}{2 i \omega A_{\rm in}(\omega)} \tilde{\rho}_{lm} (\omega) e^{-i\omega t},
\end{align}
where ${}_{-2} S_{lm}(a\omega, \theta)$ is the spin-weighted spheroidal harmonics, $R_{lm}$ is the radial Teukolsky variable, and $\tilde{\rho}_{lm}$ is
\begin{equation}
\tilde{\rho}_{lm} (\omega)= \frac{-4\omega^2}{\lambda (\lambda + 2) -12iM\omega -12 a^2 \omega^2} \int_{r_+}^{\infty} dr' \frac{\rho_{lm} (\omega, r') X_{lm}^{\rm (hom)} (\omega, r')}{A_{\rm out} (\omega)},
\end{equation}
Then we find
\begin{equation}
|\tilde{h}_{lm}| = \frac{\sqrt{1-\Gamma_{lm}}}{\omega^3}  t_{lm} = \gamma_{lm} t_{lm},
\label{ringdown_model_grey_source}
\end{equation}
where $\gamma_{lm}$ is a universal quantity that depends only on the two remnant quantities ($M,j$) and another factor $t_{lm}$ includes the source term and the spheroidal harmonics, depending on the external information like the GW source and the observation angle, respectively:
\begin{equation}
t_{lm} \equiv \left|\frac{c_0}{C} \right| |\tilde{\rho}_{lm}| |{}_{-2} S_{lm}|.
\end{equation}
The factor of $t_{lm}$ is hereinafter referred to as the {\it renormalized source term}.
The frequency dependence of the GW spectral amplitude, $|\tilde{h}_{lm}|$, is governed by $\gamma_{lm}$ at higher frequencies $\omega \gtrsim f_{lm}$ provided that $t_{lm}$ has the small dependence in $\omega$ for $\omega \gtrsim f_{lm}$. The value of the source term $t_{22}$ is shown in Figure \ref{pig_ref_grey}. One can see that $t_{22} (\omega)$ is indeed nearly constant and $\gamma_{22}$ governs the frequency dependence of the GW spectrum. It might imply that a compact object plunging into a large black hole can be regarded as an {\it instantaneous} source of GW ringdown and the associated source term can be nearly constant in the frequency domain\footnote{Remember that an instantaneous pulse like a delta function or a sharp Gaussian distribution in the time domain has a (nearly) constant distribution in the frequency domain.}.
We leave a more detailed study of our ringdown model for other harmonic modes, e.g., the sensitivity of our model for $(l,m) \neq (2,2)$ to external parameters like the orbital angular momentum, for a future work.

\subsection{Exponential decay in GW spectral amplitudes and in the greybody factors}
\label{sec_exp_decay}
\begin{figure}[t]
\centering
\includegraphics[width=0.5\linewidth]{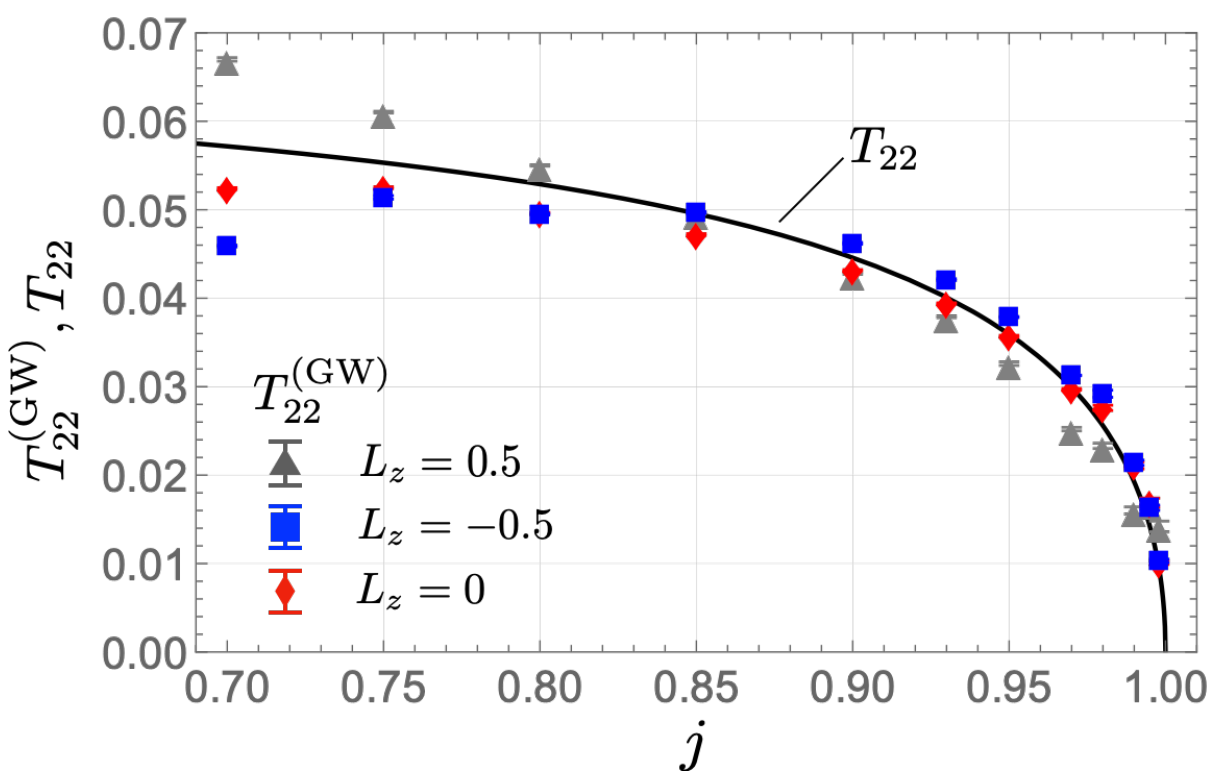}
\caption{The dots show the best fit values of $T^{\rm (GW)}_{22}$ extracted from the numerical GW waveform data with $\theta = \pi/2$. The black solid line is the best fit value of the damping frequency $T_{22}$ obtained from the numerical computation of the greybody factor.}
\label{pig_exp_damp}
\end{figure}

We find the spectral amplitude in high frequency region ($\omega \gtrsim f_{22}$) can be modeled by the greybody factor as is shown in Figures \ref{pig_comp} and \ref{pig_ref_grey}. This holds for various values of the orbital angular momentum of the plunging particle $\mu L_z$.
Fitting the Boltzmann factor $\exp[-(\omega - f_{22})/ T^{\rm (GW)}_{22}]$ to the simulated GW data\footnote{The detailed methodology of our fitting analysis is provided in Appendix \ref{app_numerical_metho}.}, we read the damping exponent of the GW spectral amplitude $T^{\rm (GW)}_{22}$ in the frequency domain. The result is shown in Figure \ref{pig_exp_damp} and the best fit values of $T_{22}^{\rm (GW)}$ (dots) is consistent with $T_{22}$ (solid line) especially for $j \gtrsim 0.8$. The value of $T_{22}$ is sensitive to the spin parameter $j$ for rapid spins, but is insensitive to $j$ for lower spins (see table \ref{table_tlm} as well).

In addition to $T_{lm}$, another quantity $f_{lm}$ is also important to model $|\tilde{h}_{lm}|$ as the exponential damping in $|\tilde{h}_{lm}|$ appears at $\omega \gtrsim f_{lm}$ (see Figures \ref{pig_spec_amp} and \ref{pig_comp}). 
In the next section, we show that the two remnant values, i.e., $M$ and $j$, can be extracted from the GW spectral amplitude by fitting $\gamma_{lm}$ characterized by $T_{lm}(M,j)$ and $f_{lm} (M,j)$. 

\subsection{estimation of the remnant quantities}
\label{sec_mismatch}
The two no-hair quantities $(M,j)$ can be extracted by fitting the function of $\gamma_{lm} (M,j) \equiv \sqrt{1-\Gamma_{lm} (M,j)}/\omega^3$ to the spectral amplitude of GW data $|\tilde{h}_{lm}|$ as the greybody factor is characterized by the two remnant quantities. The fitting parameters are $M$, $j$, and an amplitude $c_{lm}$.
Here we demonstrate the extraction of the two no-hair parameters $(M,j)$ from our clean numerical GW waveform by fitting $\gamma_{lm}$ with $(l,m)=(2,2)$. We here use the analytic model function that models $\gamma_{22}$, whose explicit form is shown in Appendix \ref{app_greybody_analytic_func}. The estimation of $(M,j)$ with noise is important to quantify the feasibility for a specific detector, and it will be studied elsewhere.

We estimate the mismatch ${\cal M}$ between the GW spectral amplitude $|\tilde{h}_{22}|$ and $c_{22} \times \gamma_{22}$ on the mass-spin space with
\begin{equation}
{\cal M} (M,a) = \left|1- \frac{\braket{|\tilde{h}_{22}||c_{22} \times \gamma_{22}}}{\sqrt{\braket{|\tilde{h}_{22}|||\tilde{h}_{22}|}\braket{c_{22} \times  \gamma_{22}|c_{22} \times \gamma_{22}}}} \right| = \left|1- \frac{\braket{|\tilde{h}_{22}||\gamma_{22}}}{\sqrt{\braket{|\tilde{h}_{22}|||\tilde{h}_{22}|}\braket{\gamma_{22}|\gamma_{22}}}} \right|,
\label{mismatch_h_g}
\end{equation}
where $\braket{a(\omega) | b(\omega)}$ is
\begin{equation}
\braket{a(\omega) | b(\omega)} = \int_{\omega_i}^{\omega_f} d\omega a(\omega) b^{\ast}(\omega).
\label{inner_prodct}
\end{equation}
Note that the mismatch ${\cal M}$ is independent of the scale $c_{22}$ and depends only on the other two fitting parameters $(M,j)$ only. This makes the fit and extraction of the remnant quantities quite simpler than the case in the multiple QN mode fitting. Also, we could avoid the overfitting issue. For the fit of multiple overtones, on the other hand, there are many fitting parameters, i.e., an amplitude and phase for each QN mode. It was pointed out \cite{Baibhav:2023clw} that the inclusion of many QN modes in the ringdown model may cause overfitting when we use a GW waveform beginning with the strain peak \cite{Baibhav:2023clw}\footnote{On the other hand, the previous work of Ref. \cite{Giesler:2019uxc} fit multiple QN modes to the numerical relativity GW waveform beginning from the strain peak. Then they reproduced the injected remnant mass and spin values. This implies that the fit of multiple QN modes may work at least when GW data has no contamination by noise.}.

\begin{figure}[h]
\centering
\includegraphics[width=0.7\linewidth]{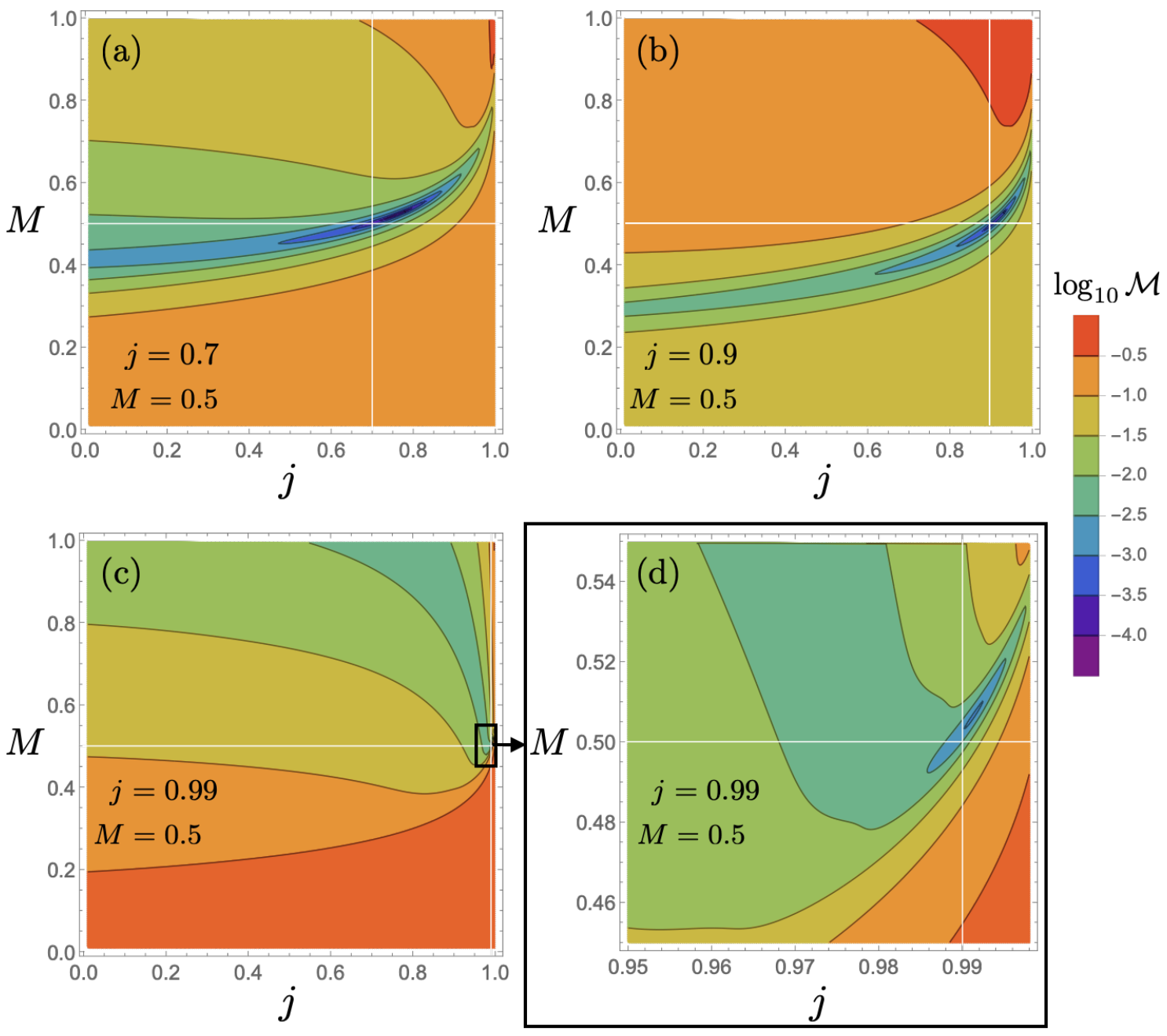}
\caption{The mismatch ${\cal M}$ between $|\tilde{h}_{22}|$ and $\gamma_{22}$ for (a) $j=0.7$, (b) $0.9$, and (c,d) $0.99$. The injected (true) values of the remnant quantities are indicated with the white solid lines. The source particle has the orbital angular momentum of $\mu L_z = 0.5 \mu$ and the observation angle is set to $\theta = \pi/2$. The frequency range used in the estimation of the mismatch is $[2M\omega_i, 2M\omega_f] = [1, 1.99]$.}
\label{pig_mass_spin}
\end{figure}

\begin{figure}[h]
\centering
\includegraphics[width=0.95\linewidth]{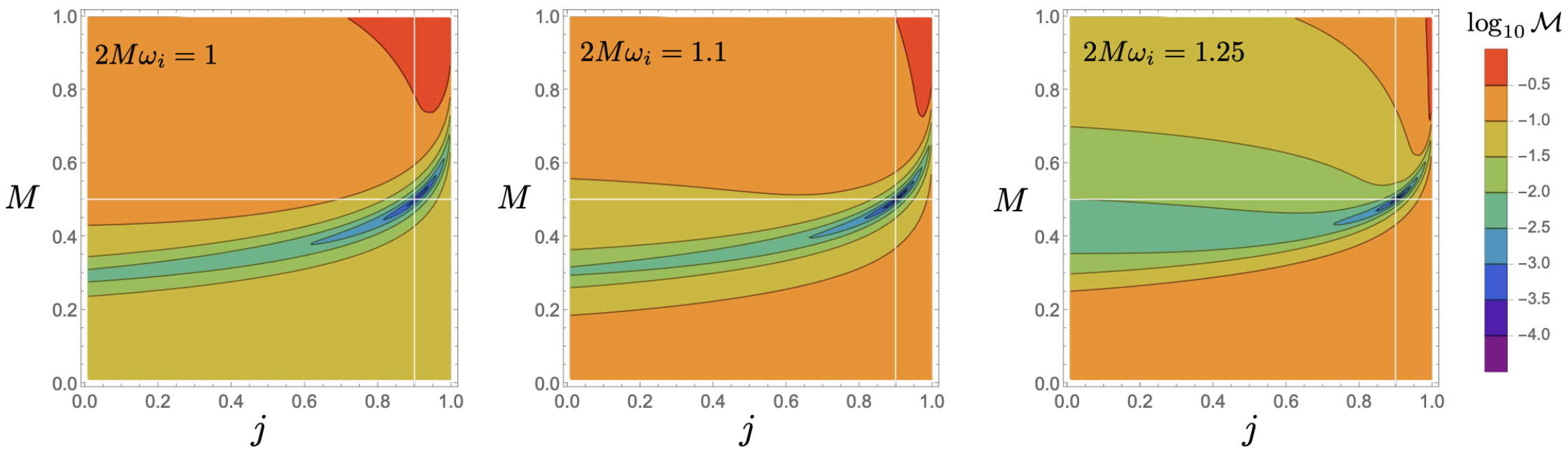}
\caption{The mismatch ${\cal M}$ between $|\tilde{h}_{22}|$ and $\gamma_{22}$ for $j=0.9$. The injected (true) values of the remnant quantities are indicated with the white solid lines. The source particle has the orbital angular momentum of $\mu L_z = 0.5 \mu$ and we set $\theta = \pi/2$. We change the frequency range of the data used in the computation of ${\cal M}$ as $2M\omega_i= 1$, $1.1$, $1.25$ and $2M \omega_f$ is fixed to $1.99$.}
\label{pig_sensitive}
\end{figure}

We estimate the mismatch ${\cal M}$ by computing the inner product (\ref{inner_prodct}) with the range of the integral of $\omega_i = 2 M$ and $\omega_f =2M \times 1.99$. Note that the $M$ in $\omega_{i/f}$ is not the true value but the fitting parameter of the black hole mass. The mismatch is computed in the mass-spin domain and the result is shown in Figure \ref{pig_mass_spin}. We find that the mass-spin estimation works well even though we here use the greybody factor without the fit of multiple QN modes. We also find that the best fit mass and spin are not sensitive to an artificial choice of the range of the data we use $\omega \in [\omega_i, \omega_f]$ as is shown in Figure \ref{pig_sensitive}. On the other hand, in the fit of QN modes, the mass-spin measurement is sensitive to the assumed start time of ringdown \cite{Giesler:2019uxc}. Although the feasibility of the extraction of the greybody factor depends on noise, combining this with the black hole spectroscopy may strengthen not only the measurability of the remnant quantities but also the precision of the test of gravity. We will come back to this in the future.

\section{Discussions}
\label{sec_conclusion}
The superposed QN modes is one of the most established model of the black hole ringdown. In this paper, we discussed another universal nature of ringdown that is described by the black hole greybody factor $\Gamma_{lm}$.
We considered how GW ringdown can be modeled by the greybody factor, which is another no-hair quantity that depends only on the mass and spin of the remnant black hole like the black hole QN modes. We found that the spectral amplitude of GW ringdown $|\tilde{h}_{lm}|$ with $(l,m)=(2,2)$ sourced by an extreme mass ratio merger can be modeled by $|\tilde{h}_{lm}| \sim c \times \gamma_{lm} (\omega)$ for $\omega \gtrsim f_{lm}$ where $\gamma_{lm} (\omega) \equiv \sqrt{1 - \Gamma_{lm}}/\omega^3$ and $c$ is an amplitude. In order for the greybody factor to be imprinted on the ringdown, the renormalized source term $t_{lm} (\omega)$, depending on the GW source, should be nearly constant with respect to $\omega$ for $\omega \gtrsim f_{lm}$. We confirmed that $t_{22} (\omega)$ satisfies the condition when GW is sourced by a compact object plunging into a massive black hole in the extreme mass ratio regime (Figure \ref{pig_ref_grey}). We may expect that this is the case as long as a particle plunging into a massive black hole can be regarded as an instantaneous source of GW ringdown. We numerically computed GW waveforms with several values of the orbital angular momentum $\mu L_z$. We then confirmed that the GW spectral amplitude is well modeled by $\gamma_{22}$, determined by the greybody factor, at higher frequencies $\omega \gtrsim f_{22}$ for various values of $|L_z| \lesssim 0.5$ (see Figures \ref{pig_comp} and \ref{pig_exp_damp}). Also, this model works well especially for a rapidly spinning remnant black hole. Indeed, the measurement of the innermost stable circular orbit of supermassive black holes (SMBHs) based on the X-ray observation puts the lower bound on the spin $j$ of SMBHs, and some of them take $j>0.9$ and can be even near extremal as $j > 0.99$ \cite{Reynolds:2019uxi,Reynolds:2020jwt}. 

As the greybody factor $\Gamma_{lm}$ is another no-hair quantity, the extraction of not only the QN modes but also the greybody factor from GW ringdown would improve the accuracy of the measurement of the remnant mass and spin and strengthens the test of gravity (Figure \ref{pig_mass_spin}).
The pros and cons in the modeling of GW ringdown with QN modes and greybody factors are summarized below.
\begin{enumerate}
    \item For the ringdown modeling with QN modes, the relevant data range in the time domain $t \gtrsim t_{\rm start}$ is difficult to identify, where $t_{\rm start}$ is the start time of ringdown. On the other hand, for the ringdown modeling with the greybody factor, the relevant data range $\omega \gtrsim f_{lm} (M,j)$ is uniquely determined once we fix the remnant quantities $M$ and $j$.
    \item Many fitting parameters are needed to extract QN modes from GW ringdown especially when several QN modes are excited simultaneously as $\sum_{n} C_{lmn} \exp [-i\omega_{lmn} t + \varphi_{lmn}]$ for a dominant angular mode of $(l,m)$. For the extraction of the greybody factor, on the other hand, the spectral amplitude of the black hole ringdown at $\omega \gtrsim f_{lm}$ is modeled by $c \times \gamma_{lm}(M,j)$. The scale $c$ is irrelevant for the minimization of the mismatch ${\cal M}$. As such, one can search the least value of ${\cal M}$ with the only two fitting parameters $(M,j)$ while avoiding the overfitting issue. It is much simpler than the QN mode fitting which involves many fitting parameters, i.e., amplitude $C_{lmn}$ and phase $\varphi_{lmn}$ for each QN mode.
    \item GW ringdown can be modeled by the superposition of QN modes regardless of the frequency-dependence of the source term. However, the modeling of ringdown with the greybody factor does not always work due to the contamination from the source term. Note that the greybody factor can be extracted only when the normalized source term $t_{lm} (\omega)$ is nearly constant in $\omega \gtrsim f_{lm}$ (Figure \ref{pig_ref_grey}).
\end{enumerate}
Given the pros and cons in the modeling of ringdown with the greybody factor and in that with QN modes, combining those two models may improve the test of the no-hair theorem and the estimation of the remnant quantities. We could also relate the excitation of overtones with the greybody factors as the residue of $\gamma_{lm}$ at QN modes can be regarded as the excitation factor, which quantifies the {\it excitability} of each QN mode \cite{Leaver:1986gd,Berti:2006wq,Zhang:2013ksa,Oshita:2021iyn}. It would be important to understand the relation between the greybody factor and excitation factor to reveal the universality in the black hole ringdown.

To further confirm the importance of the greybody factor in the modeling of GW ringdown, we have to check the detectability of the greybody factor from GW ringdown with the future detectors such as LISA. Also, it would be important to take into account some higher harmonic modes, which would affect the extraction of the greybody factor and increases the fitting parameters if higher harmonic modes are significantly excited. We will come back to these points in the future. It is interesting to note that as another different direction, the authors in Ref. \cite{Volkel:2019ahb} studied an inverse problem to read the greybody factor from quantum Hawking radiation. An interesting aspect of the greybody factor is that it can be important in both quantum and classical radiation of black holes, i.e., Hawking radiation and GW ringdown, respectively.

\begin{acknowledgements}
The author appreciate Niayesh Afshordi, Kazumasa Okabayashi, and Hidetoshi Omiya for valuable comments on this work. The author thanks Daiki Watarai for carefully reading an earlier version of this paper and for valuable comments.
The author also thanks Giulio Bonelli and Sebastian V\"{o}lkel for sharing their recent works and for helpful comments on an earlier version of this paper.
The author is supported by the Grant-in-Aid for Scientific Research (KAKENHI) project for FY2023 (23K13111).
\end{acknowledgements}

\appendix
\section{numerical methodology and accuracy}
\label{app_numerical_metho}
We numerically solve the Sasaki-Nakamura equation (\ref{SN_equation}) with the 4th Runge-Kutta method. The source term $\rho_{lm}$ for the plunging particle is numerically computed in the range of $r^{\ast}_{\rm min} \leq r^{\ast} \leq r^{\ast}_{\rm max} = 400M$. The minimum radius of the range of integral $r^{\ast}_{\rm min}$ is set to
\begin{align}
r^{\ast}_{\rm min} =
\begin{cases}
-40M, \ &\text{for} \ j \leq 0.9,\\
-50M, \ &\text{for} \ 0.9 < j < 0.97,\\
-80M, \ &\text{for} \ 0.97 \leq j < 0.99,\\
-100M, \ &\text{for} \ j = 0.99,\\
-120M, \ &\text{for} \ j = 0.995,\\
-160M, \ &\text{for} \ j = 0.998.
\end{cases}
\end{align}
For the Sasaki-Nakamura equation, the exponential tail of the potential $U_{lm}$ near the horizon becomes long range as $j \to 1$. As such, $r^{\ast}_{\rm min}$ should be a larger negative value for rapid spins so that one can impose the boundary condition of $e^{-i k_{\rm H} r^{\ast}}$ at the end point of $r^{\ast} = r^{\ast}_{\rm min}$.
The numerical integration of the Sasaki-Nakamura equation is done in the range of $r^{\ast}_{\rm min} \leq r^{\ast} \leq r^{\ast}_{\rm SN max} = 300 + 30 / \omega$ for each frequency mode of $\omega$. 
\begin{figure}[t]
\centering
\includegraphics[width=0.55\linewidth]{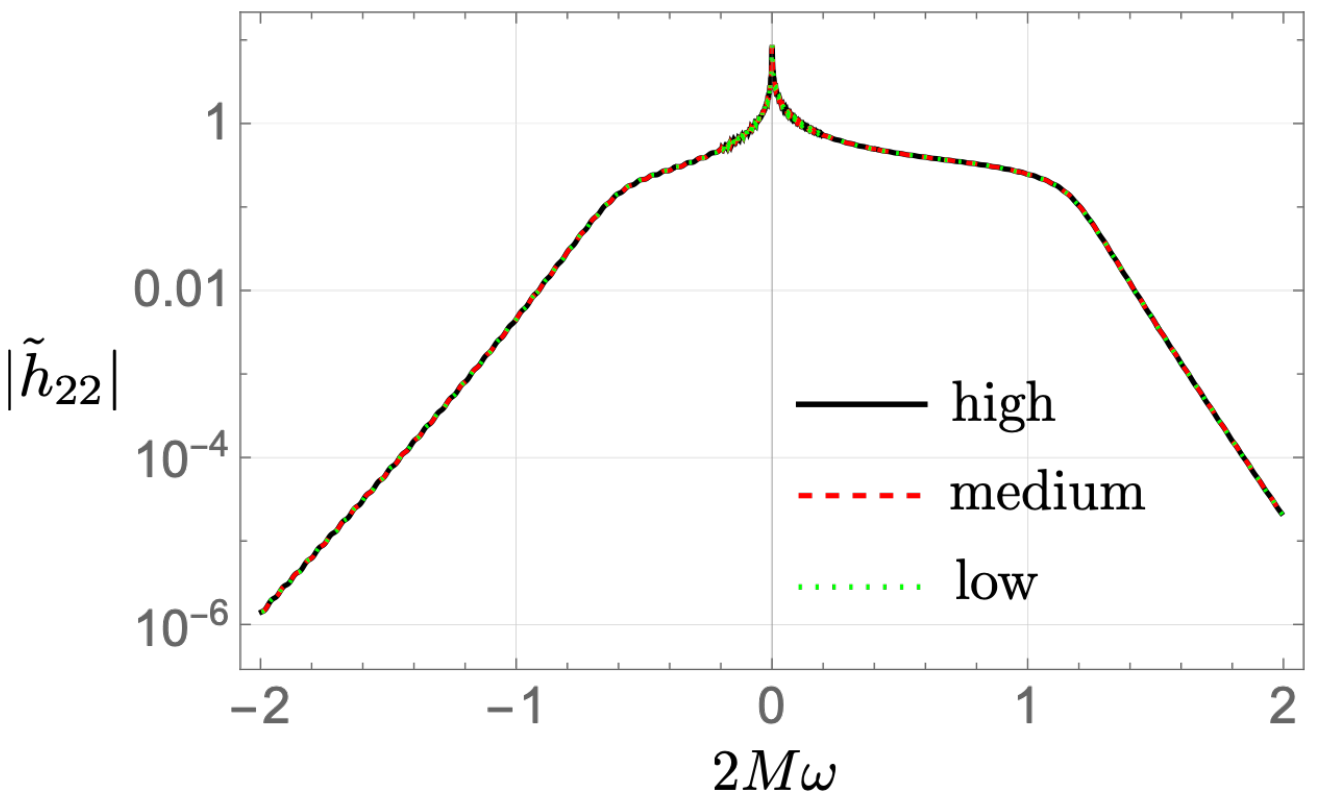}
\caption{GW spectral amplitude for $j=0.8$, $(l,m)=(2,2)$, $L_z = 0.5$ and $\theta = \pi/2$. We compute it for high, medium, and low resolutions with $(N_{\rm source}, N_{\rm SN}) = (10^4, 2 \times 10^5)$, $(5 \times 10^3, 10^5)$, and $(2.5 \times 10^3, 5 \times 10^4)$, respectively.}
\label{pig_resolution}
\end{figure}

The source term $\rho_{lm} (\omega, r^{\ast})$ is obtained in the resolution of $\Delta r^{\ast} = (r^{\ast}_{\rm max} - r^{\ast}_{\rm min})/N_{\rm source}$ with $N_{\rm source} = 5000$. The Sasaki-Nakamura equation is integrated with the step size of $\Delta r^{\ast} = (r^{\ast}_{\rm SN max} - r^{\ast}_{\rm min})/N_{\rm SN}$ with $N_{\rm SN} = 10^5$. The greybody factor is computed by reading the asymptotic amplitude at $r^{\ast} = r^{\ast}_{\rm SN max}$ by using the Wronskian.
We checked that our resolution is high enough to obtain high-accuracy GW waveform and greybody factor (see Figure \ref{pig_resolution}).

The damping frequency $T_{lm}$ in $1-\Gamma_{lm} (\omega)$ and $T_{lm}^{\rm (GW)}$ in the GW spectral amplitude $|\tilde{h}_{lm} (\omega)|$ are extracted at higher frequencies $\omega \gtrsim f_{lm}$ by using a Mathematica's function \texttt{NonlinearModelFit} for the log-scaled data, $\log \Gamma_{lm}$ and $\log (|\tilde{h}_{lm}|^2 \omega^6)$, with the fitting function of
\begin{equation}
B (\omega - \omega_i) + \log A,
\end{equation}
where $A$ and $B$ are the fitting parameters and $B$ is associated with $T_{lm}$ or $T^{\rm (GW)}_{lm}$. 
The results for $(l,m) = (2,2)$ are shown in Table \ref{table_tlm} and Figure \ref{pig_exp_damp}. The extraction of $T_{22}$ is done by fitting the Boltzmann factor $e^{-(\omega - f_{22})/T_{22}}$ to the data in the frequency range of $\omega_i \leq \omega \leq \omega_f = 1.99/(2M)$ with
\begin{align}
\omega_i =
\begin{cases}
1.20 \times f_{22}, \ &\text{for} \ 0.001 \leq j \leq 0.75,\\
1.10 \times f_{22}, \ &\text{for} \ 0.8 \leq j \leq 0.9,\\
1.05 \times f_{22}, \ &\text{for} \ 0.93 \leq j \leq 0.99,\\
1.02 \times f_{22}, \ &\text{for} \ j=0.995,\\
1.00 \times f_{22}, \ &\text{for} \ j= 0.998.
\end{cases}
\label{greybody_reading}
\end{align}
For the extraction of $T_{22}^{\rm (GW)}$, we fit the Boltzmann factor to the numerical data in the range of $\omega_i \leq \omega \leq \omega_f$ with
\begin{align}
\omega_i =
\begin{cases}
1.20 \times f_{22}, \ &\text{for} \ 0.001 \leq j \leq 0.75,\\
1.10 \times f_{22}, \ &\text{for} \ 0.8 \leq j \leq 0.9,\\
1.05 \times f_{22}, \ &\text{for} \ 0.93 \leq j < 0.98,\\
1.00 \times f_{22}, \ &\text{for} \ 0.98 \leq j \leq 0.998,
\end{cases}
\end{align}
and $\omega_f$ is set to a value at which $(\omega_f^3|\tilde{h}_{22}(\omega_f)|^2)/(\omega_i^3|\tilde{h}_{22}(\omega_i)|^2) \simeq 0.01$. Also, the best fit value and error of $T_{22}^{\rm (GW)}$ in Figure \ref{pig_exp_damp} was estimated by Mathematica's commands \texttt{BestFitParameters} and \texttt{ParameterErrors} in a Mathematica's function of \texttt{NonlinearModelFit}.
\begin{figure}[h]
\centering
\includegraphics[width=0.9\linewidth]{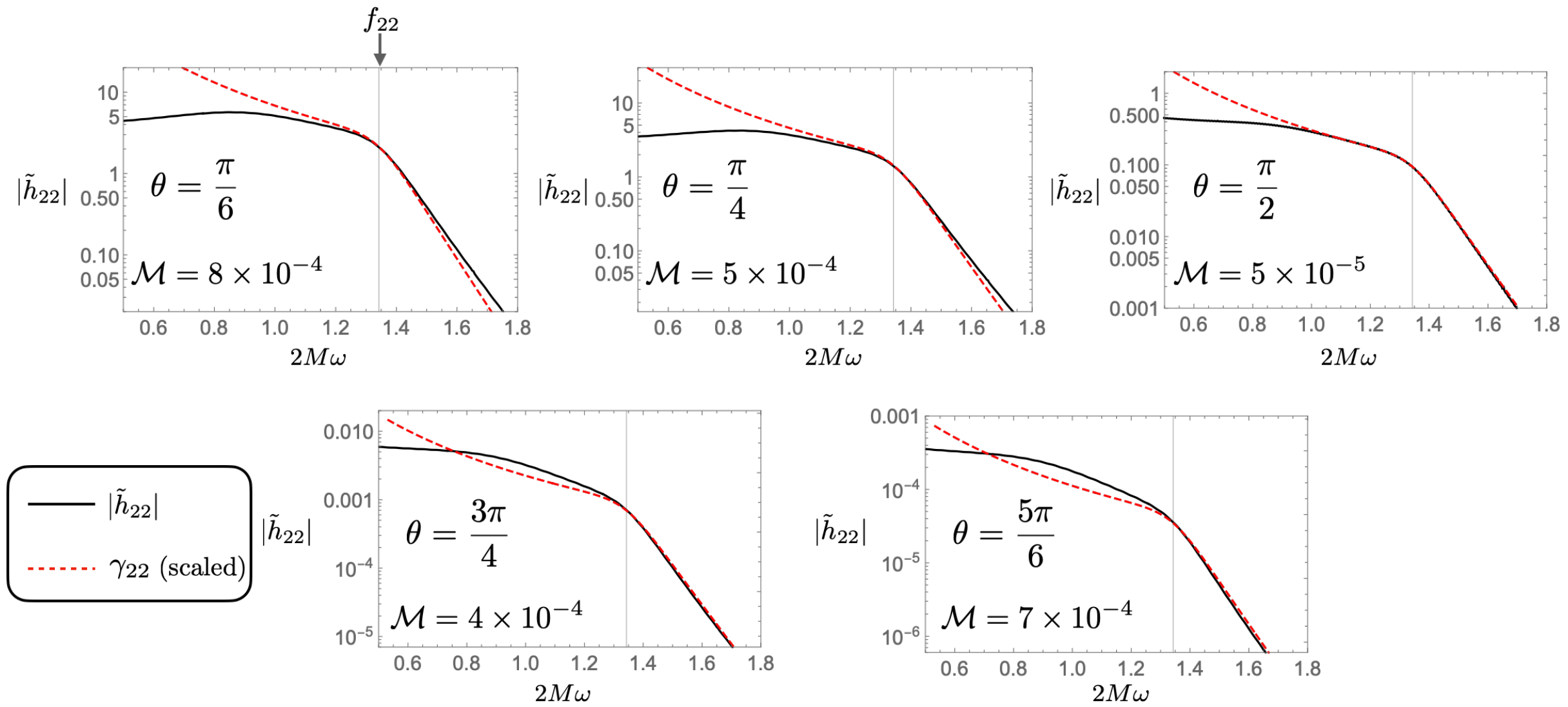}
\caption{Comparison of the spectral amplitude of GW $|\tilde{h}_{22}|$ (black solid) and $\gamma_{22}$ (red dashed). We set $j=0.9$ and $L_z=0.5$. We also set the observation angle as $\theta=\pi/6$, $\pi/4$, $\pi/2$, $3\pi/4$, and $5\pi/6$. The mismatch ${\cal M}$ is evaluated for data in $f_{22} \leq \omega \leq 1.99/(2M)$.}
\label{pig_observation_angle}
\end{figure}

\section{greybody factor in GW ringdown and observation angle $\theta$}
\label{app_theta}
Our ringdown modeling for a harmonic mode $(l,m)$ is given by the product of $\gamma_{lm}$ and the renormalized source term (\ref{ringdown_model_grey_source}). The renormalized source term is determined by a source of GW emission and the observation angle as it includes the spin-weighted spheroidal harmonics ${}_{-2}S_{lm} (a\omega, \theta)$. We confirmed that the ringdown modeling with the greybody factor works for a wide range of the observation angle $\theta$.
Indeed, the mismatch ${\cal M}$ defined in (\ref{mismatch_h_g}) is less than $10^{-3}$ at least for $\pi/6 \leq \theta \leq 5\pi/6$ as is shown in Figure \ref{pig_observation_angle}. The mismatch ${\cal M}$ is evaluated for data in $f_{22} \leq \omega \leq 1.99/(2M)$.

\section{analytic model of the greybody factor}
\label{app_greybody_analytic_func}
\begin{figure}[t]
\centering
\includegraphics[width=1\linewidth]{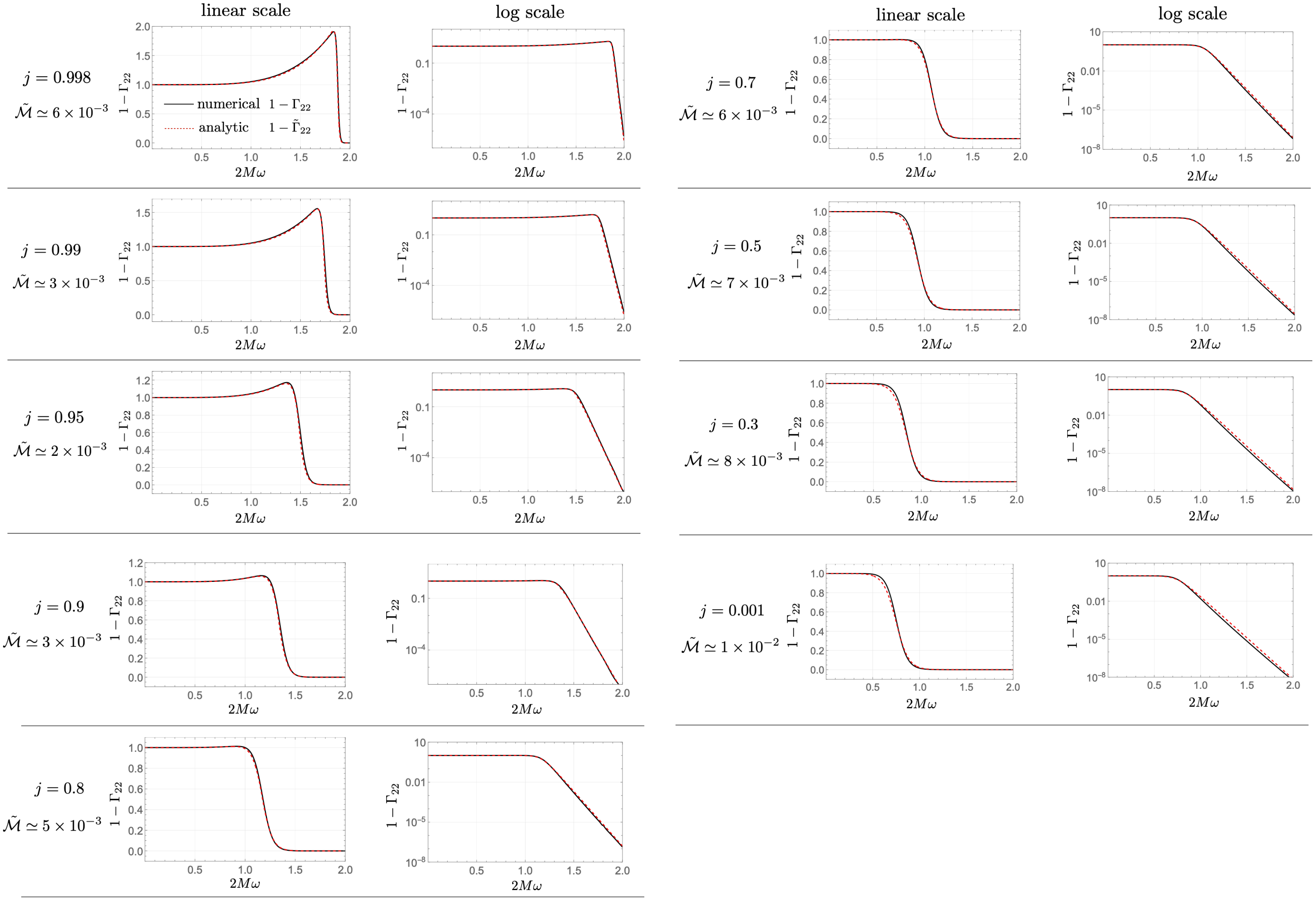}
\caption{Reflectivity ${\cal R}_{22} = 1-\Gamma_{22}$ is shown in the linear and log scale for several spin parameters in $0.001 \leq j \leq 0.998$. The mismatch $\tilde{\cal M}$ between the reflectivity obtained by our numerical computation $1-\Gamma_{22}$ and the one modeled by the analytic function in (\ref{reflectivity_model}).}
\label{pig_grey_model}
\end{figure}

Our proposal in this paper is that the greybody factor $\Gamma_{lm}$ is imprinted on the spectral amplitude of GW ringdown with the form of
\begin{equation}
\gamma_{lm} = \sqrt{1- \Gamma_{lm}}/\omega^3.
\end{equation}
As the greybody factor is a universal quantity which depends only on the remnant mass and spin like the black hole QN modes, the extraction of the greybody factor from the signal is applicable to test the no-hair theorem and the measurement of the remnant mass and spin. To demonstrate that in Sec. \ref{sec_mismatch}, we compute the mismatch ${\cal M}$ between GW spectral amplitude and the function $\gamma_{lm}$. As the computation of the greybody factor involves the numerical integration of the Sasaki-Nakamura equation in our approach, we shorten the computation time of ${\cal M}$ by using an analytic model function $\tilde{\Gamma}_{22}$ that models the greybody factor $\Gamma_{22}$ for $\omega > 0$\footnote{Another fitting function of the reflectivity for $0.6 < j < 0.8$ is provided in Ref. \cite{Nakano:2017fvh}.}:
\begin{equation}
1-\tilde{\Gamma}_{22} (\omega) = \frac{1 + a_1 Z[-2,2,2,M,j,\omega] (1-\tanh{[(\omega - f_{22})/a_2]})}{(1 + \exp[(\omega - f_{22})/T_{22}])} \ \text{for} \ \omega > 0,
\label{reflectivity_model}
\end{equation}
where $a_1 = 0.325$, $a_2=0.02$, and
\begin{align}
T_{22} &\simeq 0.223 \sqrt{1-j} -0.33 (1-j) + 0.249 (1-j)^{3/2} - 0.0748 (1-j)^2,\\
f_{22} &\simeq 2 - 2.85 \sqrt{1-j} + 3.01 (1-j) -2.01 (1-j)^{3/2} + 0.597 (1-j)^2,\\
Z[s,l,m,M,j,\omega] &\equiv 4 m \Omega_{\rm H} \frac{r_+}{\sqrt{1-j^2}} \left( \frac{(l-s)! (l+s)!}{(2l)! (2l+1)!!} \right)^2 [2 M \omega (1-j^2)]^{2l+1} \prod_{k=1}^l \left( 1+ \frac{4}{k^2} \left(m \Omega_{\rm H} \frac{r_+}{\sqrt{1-j^2}} \right)^2 \right),
\end{align}
where $s$ is the spin of the relevant field, e.g., $|s|=2$ for gravitational field.
This fitting model matches with the exact greybody factor within $\tilde{{\cal M}} \lesssim 0.01$ as is shown in Figure \ref{pig_grey_model}. This fitting function is applicable to the broad range of spin parameter $0.001 \leq j \leq 0.998$ as is partially shown in the Figure.

\end{document}